\documentclass[aps,prl,reprint,a4paper]{revtex4-1}

\usepackage{amsmath}
\usepackage{bm}
\usepackage{siunitx}
\usepackage{graphicx}

\newcommand{\br}{{\bm r}}
\newcommand{\kT}{k_\mathrm{B}T}
\newcommand{\so}
   {\mathrel{\rlap{\raise1pt\hbox{$>$}}{\lower4pt\hbox{$\sim$}}}}
\newcommand{\io}
   {\mathrel{\rlap{\raise1pt\hbox{$<$}}{\lower4pt\hbox{$\sim$}}}}

\begin{document}

\title{Membrane protein clustering from tension and multibody\\interactions}

\author{Jean-Baptiste Fournier}
\affiliation{Universit\'e Paris Cit\'e, CNRS, Laboratoire Mati\`ere et Syst\`emes Complexes (MSC), F-75013 Paris, France}

\begin{abstract}
The point-curvature model for membrane protein inclusions is shown to capture {multibody interactions very well. Using this model, we find} that the interplay between membrane tension and multibody interactions results in a collective attraction of oppositely curved inclusions tending to form antiferromagnetic structures with a square lattice. This attraction can produce a phase separation between curved and non-curved proteins, resulting in the clustering of curved proteins. We also show that the many-body repulsion between identical curved proteins is enhanced by membrane tension. This can lead to the dissolution of clusters stabilized by short-range forces when the tension is increased. These new phenomena are biologically relevant and could be investigated experimentally
\end{abstract}

\maketitle

\begin{figure}[b]
\centerline{\includegraphics[width=.7\columnwidth]{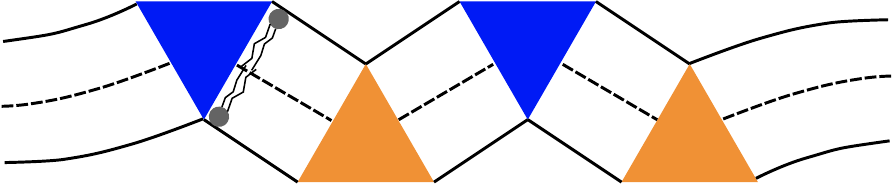}}
\caption{One-dimensional sketch of curvature-inducing membrane proteins arranged antiferromagnetically.}
\label{fig:sketch}
\end{figure}

Biological membranes are bilayers of amphiphilic lipid molecules surrounded by an aqueous solvent, which constitute the wall of our cells~\cite{Lodish_book}. Their elasticity is governed by bending rigidity and lateral tension~\cite{Helfrich:1973}. Curvature-inducing integral proteins with conical shape deform membranes on a large scale and undergo long-range interactions mediated by the deformation of the membrane~\cite{Goulian:1993,Kim:1998,Dommersnes:1999,Dommersnes:2002,Reynwar:2007,Reynwar:2011,Yolcu:2012,Yolcu:2014,Fournier:2015,Vahid:2016,Kohyama:2019,Galatola:2023} (fig.~\ref{fig:sketch}).
Membrane proteins interact via a variety of other mechanisms, such as short-range interactions~\cite{Sieber:2007,Destainville:2008}, the interplay between shape and hydrophobic mismatch~\cite{Fournier:1998,Morozova:2012}, interaction with lipid recruitment, depletion, line tension and with protein scaffolds~\cite{Goyette:2017,Johannes:2018}, cooperativity with receptor-ligand binding~\cite{Li:2021} and through membrane reshaping~\cite{Mondal:2023}. In all these mechanisms, however, the direct effect of membrane tension is largely ignored.

In a tensionless membrane with bending rigidity $\kappa$, two curvature-inducing inclusions with circular cross-section, imposing local curvatures $c_1$ and $c_2$ of any sign, are known to repel each other with an asymptotic energy proportional to $\kappa(c_1^2+c_2^2)(a/r)^4$, where  $a$ is the radius of the inclusions and $r$ their separation. This was first shown in the framework of the disk with detachment angle (DDA) model introduced by Goulian et al.~\cite{Goulian:1993}. The corresponding interaction has been calculated as a power series up to order $1/r^{14}$~\cite{Fournier:2015} and computed numerically at very short separations~\cite{Galatola:2023}.
It is valid in the small deformation limit; in the nonlinear regime, numerical calculations indicate that identical inclusion may attract each other at very small separations~\cite{Reynwar:2007,Reynwar:2011}. Note that inclusions with anisotropic cross-sections, or inclusions inducing anisotropic curvatures, undergo a stronger interaction $\sim r^{-2}$~\cite{Park:1996,Dommersnes:1999,Galatola:2023}. 

In the presence of a significant lateral tension $\sigma$, the interaction between curvature-inducing proteins relaxes exponentially over the correlation length $\xi=\sqrt{\kappa/\sigma}$, resulting however in increased repulsion at short separations for inclusions of the same curvature sign, and short-range attraction with minimum-energy separation for inclusions of opposite curvature signs~\cite{Weikl:1998}. In addition to these deformation-mediated interactions, membrane inclusions also undergo fluctuation-induced interactions (Casimir-like forces), but these are negligible for proteins inducing large local curvatures~\cite{Goulian:1993,Dommersnes:1999,Lin:2011}.

Remarkably, {within their total many-body energy}, membrane inclusions undergo multibody interactions as prominent as pairwise interactions~\cite{Kim:1998,Galatola:2023}. Large clusters of inclusions were originally studied in Ref.~\cite{Kim:1998}, however with a method based on the energy required to insert a protein into a curved background. This method correctly captures pairwise and triplet interactions, but neglects higher order multibody interactions~\cite{Yolcu:2014}. Exact results on multibody interactions were obtained within the DDA model very recently~\cite{Galatola:2023}. {In the presence of tension, the dominant three-body interaction was derived in Ref.~\cite{Weitz:2013}, but little is known about the overall importance of multibody interactions in equilibrium clusters.}

{In this Letter, we apply the pointlike curvature constraint model introduced in Refs.~\cite{Dommersnes:1999,Dommersnes:2002,Weitz:2013} to proteins embedded in membranes under tension}. This method is simpler than the traditional DDA method and allows quite a large number of inclusions to be treated exactly within its framework. We show (i) that the pointlike model captures the asymptotic triplet and quadruplet interactions in exactly the same way as the DDA model in the absence of tension, and (ii) we verify that it captures correctly the pairwise interaction in the presence of tension. We then apply the pointlike method to sets of curvature-inducing protein inclusions, showing that the enhancement of the many-body interaction due to membrane tension can lead to novel clustering behaviors.

In the small deformation limit, which we adopt, the membrane Hamiltonian reads~\cite{Helfrich:1973}
\begin{align}
H=\int d^2r\left[\frac\kappa2\left(\bm\nabla^2 h\right)^2+\frac\sigma2\left(\bm\nabla h\right)^2\right],
\label{eq:H}
\end{align}
where $h(\br)$ denotes the height of the membrane above a reference plane parametrized by $\br=(x,y)$. The first term describes the bending energy of the membrane and the second term describes the cost of a deformation when the membrane is subjected to lateral tension. Bare membrane correlations are given by the Green's function of $\mathcal L=\bm\nabla^4-\xi^{-2}\bm\nabla^2$, i.e., $G(\br)=-(2\pi)^{-1}\xi^2[K_0(r/\xi)+\ln(r/\xi)]$, with $K_0$ the zeroth order Bessel function.

We consider a set of $N$ protein inclusions, each locally imposing an isotropic curvature $c_i$ at $\br=\br_i$, i.e., $\partial^2_xh(\bm r_i)=\partial^2_yh(\bm r_i)=c_i$ and $\partial_x\partial_yh(\bm r_i)=0$. {Note that anisotropic proteins could be dealt with using anisotropic curvature tensors~\cite{Dommersnes:1999}}. It was shown in Refs.~\cite{Dommersnes:1999,Dommersnes:2002} that its  many-body energy is given by
\begin{align}
H=\frac\kappa2\mathsf{C}^t\mathsf M^{-1}\mathsf C.
\label{eq:H_PCC}
\end{align}
with $\mathsf C^t=(c_1,c_1,0,\ldots,c_N,c_N,0)$, corresponding, in order to $\partial^2_xh$, $\partial^2_yh$ and $\partial_x\partial_yh$ for each inclusion, and
\begin{align}
\mathsf M=\begin{pmatrix}
m_{11}&\ldots &m_{1N}\cr
\vdots &\ddots &\vdots\cr
m_{N1}&\ldots &m_{NN}
\end{pmatrix},
~m_{ij}=
\begin{pmatrix}
G_{ij}^{(4)} & G_{ij}^{(2)} &
G_{ij}^{(3)}
\cr
G_{ij}^{(2)} & G_{ij}^{(0)} &
G_{ij}^{(1)}
\cr
G_{ij}^{(3)} & G_{ij}^{(1)} &
G_{ij}^{(2)}
\end{pmatrix},
\label{eq:matrix}
\end{align}
where $G_{ij}^{(4)}=\partial^4_xG(\br_j-\bm r_i)$, $G_{ij}^{(3)}=\partial^3_x\partial_y(\br_j-\bm r_i)$, etc., are the fourth derivatives of the Green's function, the superscript denoting how many derivatives are taken with respect to~$x$ (see Appendix). The diagonal components $m_{ii}$ diverge and must be 
regularized in reciprocal space by an upper-wavevector cutoff $q_\text{max}$ giving an effective size to the pointlike inclusion (see Appendix). We take $q_\text{max}=2/a$, so that the pointlike model matches asymptotically the DDA model for a disk of radius $a$. Then, $G_{ii}^{(4)}=G_{ii}^{(0)}=3g$, $G_{ii}^{(2)}=g$ and $G_{ii}^{(3)}=G_{ii}^{(1)}=0$, with
$g=[4\xi^2/a^2-\ln(1+4\xi^2/a^2]/(32\pi\xi^2)$. In the following, $E$ represents the true many-body interaction energy, obtained by subtracting the energies of the isolated inclusions, i.e., $E=H-\sum_iH_i$.

\begin{figure}
\centerline{\includegraphics[width=.8\columnwidth]{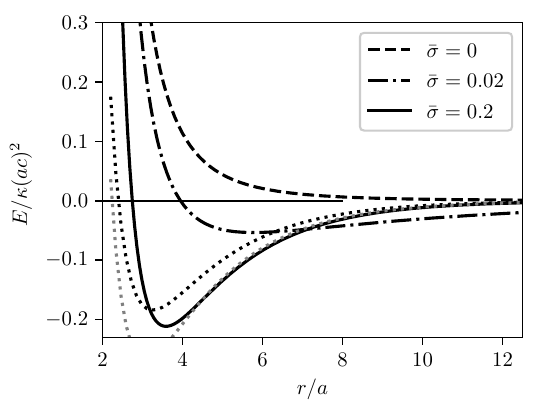}}
\vspace{-.4cm}
\caption{Pairwise interaction energy $E$ of two protein inclusions with opposite curvatures $\pm c$, plotted against their separation $r$, for different membrane tensions $\sigma=\bar\sigma\kappa/a^2$. These results are obtained from the point curvature model, given by Eq.~\eqref{eq:H_PCC}. The black dotted line shows the approximation~\eqref{eq:Weikl}, originally obtained under the standard DDA model~\cite{Weikl:1998}, for comparison. {Note that for $\bar\sigma=0.2$, as $\xi/a=1/\sqrt{\bar\sigma}\simeq2.2$, the whole curve is in the  $r\so\xi$ regime.}}
\label{fig:pairwise}
\end{figure}

\section{Paiswise interactions}
For two proteins, in the absence of tension, Eq.~\eqref{eq:H_PCC} gives asymptotically
\begin{align}
E\simeq4\pi\kappa a^2\left(c_1^2+c_2^2\right)\left(\frac ar\right)^4,
\label{eq:Goulian}
\end{align}
which coincides with the result of  Ref.~\cite{Goulian:1993} obtained within the DDA model. In the presence of tension, for $a\ll\xi$ and $r<\xi$, Eq.~\eqref{eq:H_PCC} gives
\begin{align}
E\simeq\pi\sigma a^4\left[
2c_1c_2K_0\!\left(\frac r\xi\right)+
\left(c_1^2+c_2^2\right)
\left(\frac a\xi\right)^2\!
K_2^2\!\left(\frac r\xi\right)
\right],
\label{eq:Weikl}
\end{align}
where the $K_i$'s are Bessel functions, which matches again the result of Ref.~\cite{Weikl:1998} obtained within the DDA model. {Note that for $r\so\xi$, the point model gives two other terms $\mathcal O(a/\xi)^6$~\cite{Weitz:2013}, yielding the gray dotted curve in Fig.~\ref{fig:pairwise}.} The interaction is attractive for oppositely curved inclusions at large tension, as shown in fig.~\ref{fig:pairwise}. The point curvature model and the DDA model thus give very similar results even in the presence of tension. Note however, that they differ qualitatively when $r$ is too small, as they encode differently the shape of the inclusion~\cite{Yolcu:2014,Fournier:2015}. In the following, to get reliable results, we shall restrict the approach of two inclusions by using a hard-core diameter equal to 3--4 times the cutoff~$a$.

\begin{figure}
\centerline{\includegraphics[width=.5\columnwidth]{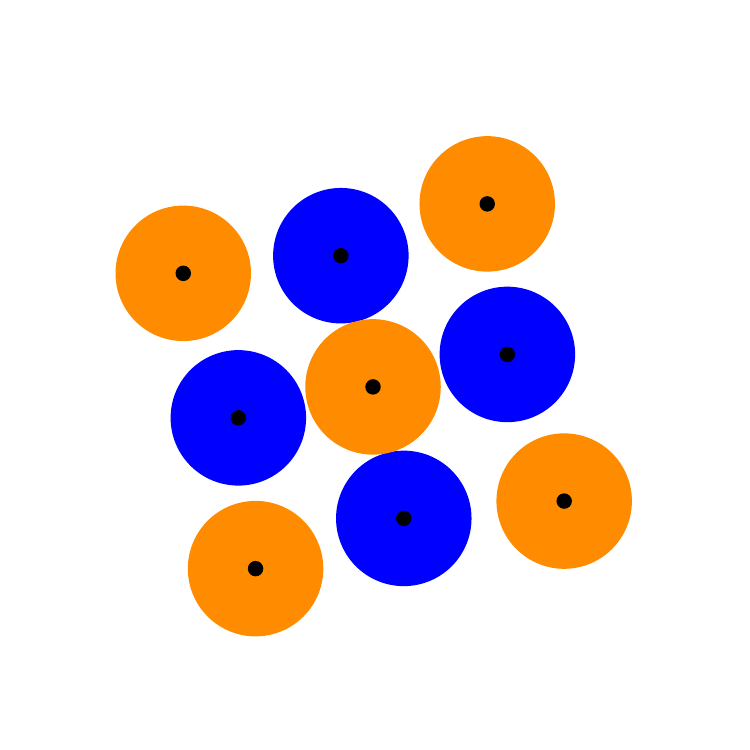}}
\vspace{-.4cm}
\caption{Minimum energy configuration of a cluster of $N=9$ inclusions with hard core radius $r_\text{hc}=2a$ (radius to scale) and $\bar\sigma=0.02$, obtained by simulated annealing. Blue and orange represent particles of opposite curvatures.}
\label{fig:siman}
\end{figure}

\section{Orders of magnitude}
For the voltage-dependent K$^+$ channel KvaP, the spontaneous curvature $c\simeq(\SI{25}{nm})^{-1}$ was measured in Ref.~\cite{Aimon:2014}. With $a\simeq\SI{5}{nm}$~\cite{Fribourg:2014}, this yields $c\simeq0.2\,a^{-1}$. For the bacterial multidrug ATP-binding cassette BmrA, measured values are $a\simeq\SI{4}{nm}$ and $c\simeq(\SI{7}{nm})^{-1}$ in the ATP-blocked configuration~\cite{Bassereau:2023}. This yields the larger value $c\simeq0.57\,a^{-1}$. Since we are interested in strongly curved inclusions in biological membranes, we will assume in the following
\begin{align}
a=\SI{4}{nm},\quad c=0.5\,a^{-1},\quad \kappa=30\,\kT.
\label{eq:params}
\end{align}
The dimensionless tension $\bar\sigma=0.2$ in fig.~\ref{fig:pairwise} then corresponds to $\sigma=\SI{1.5e-3}{J/m^2}$, which is large but smaller than the typical lysis tension, and the minimum energy corresponds to $E_\text{min}\simeq-1.6\,\kT$ and $r_\text{min}\simeq\SI{15}{nm}$.

\section{Multibody interactions}
To support the results of this paper on many-body interactions, we show here that the point model and the DDA model give the same exact asymptotic interaction for small clusters, in situations where three- and four-body interactions are present in the absence of tension. We have calculated with Eq.~\eqref{eq:H_PCC} the total (i.e., many-body) energy $E_\text{tri}$ of triangular and $E_\text{sqr}$ of square clusters of $N=3,4$ identical inclusions of curvature $c$. The positions of the inclusions are given in polar coordinates by $\theta_k=2\pi k/N$ and $r_k=R$, with $0\le k<N$. To leading order in $1/R$, we obtain
\begin{align}
E_\text{tri}\simeq\frac43\pi\kappa a^2c^2\left(\frac aR\right)^4,
\quad
E_\text{sqr}\simeq\pi\kappa a^2c^2\left(\frac aR\right)^4.
\end{align}
The same quantities were calculated within the DDA model in Ref.~\cite{Galatola:2023} yielding the same exact results. Detailed analysis (to be published) shows that the pairwise components of the interactions are repulsive, while the three-body and four-body components are attractive and comparable to the total many-body energy.

\begin{figure}
\centerline{\includegraphics[width=.8\columnwidth]{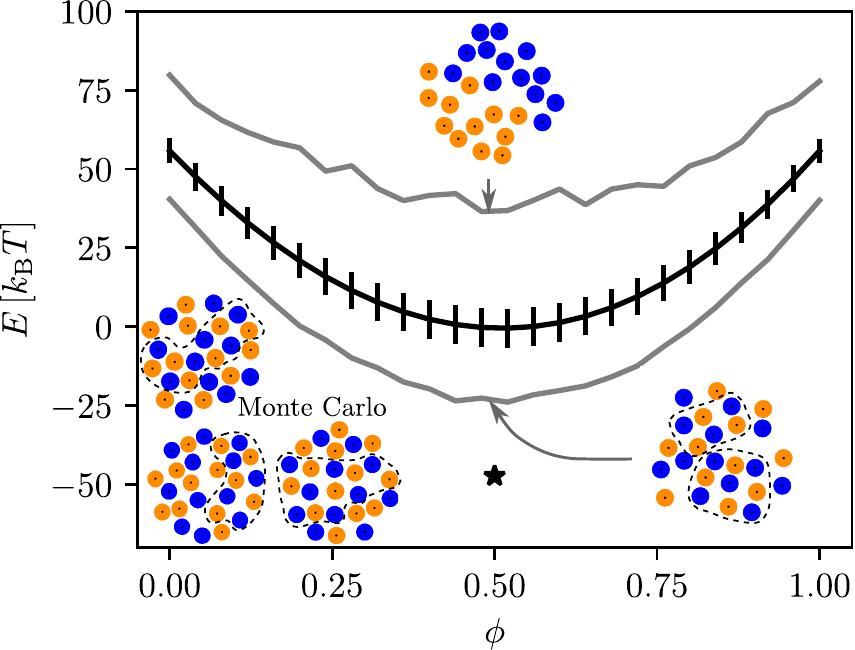}}
\caption{Many-body energy landscape. (i) Monte Carlo simulation snapshots of $N=25$ inclusions with hard-core radius $2a$ and curvatures $\pm c$, with a composition $\phi=50\%$, confined in a disk of radius $R$, showing disordered antiferromagnetic arrangements (lower left corner), and (ii) Energy distribution of $P=10^7$ random clusters of $N=25$ inclusion with curvatures $\pm c$, confined in a disk of radius $R$, as a function of  composition $\phi$: the black line with error bars gives the average energy and  standard deviation; the top and bottom gray lines are the maximum and minimum energies, with corresponding typical clusters at $\phi=0.5$; the star indicates the energy of the antiferromagnetic cristalline structure with a square lattice of spacing $d=4a$ (fig.~\ref{fig:tension_on_crystal}). Parameters: $R=15a$, $c=0.5\,a^{-1}$, $\kappa=30\,\kT$ and $\bar\sigma=\sigma a^2/\kappa=0.2$.}
\label{fig:smiley}
\end{figure}

\section{Mixtures of oppositely curved inclusions}
Since inclusions with opposite curvatures attract each other in tense membranes, we studied  clusters of opposite $\pm c$ curvatures. First, using simulated annealing, we determined the minimum energy configuration of a small cluster of $N=9$ inclusions within the point-like model, using the tension $\bar\sigma=0.2$ giving a well-defined minimum in pairwise energy. In order to sample the phase space efficiently, we allowed inclusion exchanges in addition to displacements. Whatever the initial condition, we obtained the crystalline structure with antiferromagnetic order shown in fig.~\ref{fig:siman}. Note that the inclusions in the corner are slightly off-lattice.

Monte Carlo simulations of large systems of inclusions ($N\gg1$) are not feasible due to the $3N\times3N$ size of the $\mathsf M$ matrix, which must be inverted, and the complexity of its elements (see Appendix). We therefore simulated, using the Metropolis algorithm, medium-sized clusters of $N=25$ inclusions of $\phi=N^+/N=50\%$ composition, $N^+$ being the number of inclusions with curvature $+c$, with hard-core radius $r_\text{hc}=2a$, for the reduced tension $\bar\sigma=0.2$. Typical clusters show liquid behavior with disordered antiferromagnetic arrangements  (fig.~\ref{fig:smiley}).  

\begin{figure}
\centerline{\includegraphics[width=.9\columnwidth]{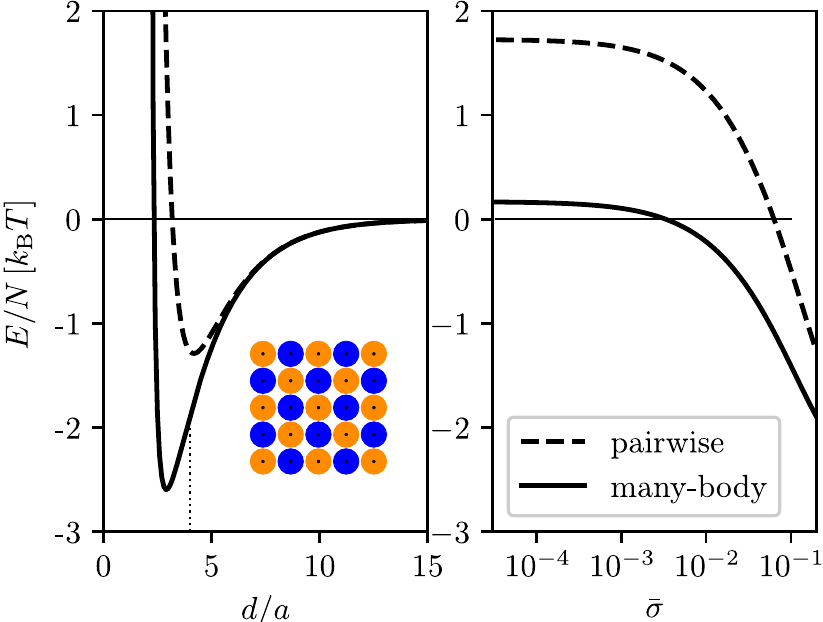}}
\caption{Many-body energy per particle (solid line) in an antiferromagnetic crystalline cluster with square lattice of spacing $d$, made of $N=25$ inclusions with curvatures $\pm c$ (see inset), shown as a function of $d$ for $\bar\sigma=0.2$ (left) and as a function of $\bar\sigma=\sigma a^2/\kappa$ for $d=4a$ (right). The dashed lines show the energy calculated from the sum of the pairwise interactions. Parameters: $c=0.5\,a^{-1}$ and $\kappa=30\,\kT$.
}
\label{fig:tension_on_crystal}
\end{figure}

Next, to get an idea of the energy landscape, we randomly generated {(with a uniform distribution)} a large number $P$ of configurations, also varying  the $\phi$ composition. We determined both the extremal energy configurations and the standard deviation of the energy (fig.~\ref{fig:smiley}). Not surprisingly, we found that mixed clusters with $0.25\io\phi\io0.75$ have a negative interaction energy $E$, when they form local antiferromagnetic cristalline structures with a square lattice (fig.~\ref{fig:smiley}, bottom right). The energy of a  perfect $5\times5$ cristal (fig.~\ref{fig:smiley}, bottom left), was significantly lower, with an energy per particle of $-1.9\,\kT$.

As shown in fig.~\ref{fig:tension_on_crystal}, this antiferromagnetic cristalline structure is stabilized by the interplay of membrane tension and multibody interactions. Note that the energy of any  configuration is proportional to $\kappa c^2$, at fixed $\sigma a^2/\kappa$, which allows the energies in figs.~\ref{fig:smiley} and \ref{fig:tension_on_crystal} to be calculated for other parameter sets.

\section{Mixtures with cylindrical proteins}
\begin{figure}
\centerline{\includegraphics[width=.75\columnwidth]{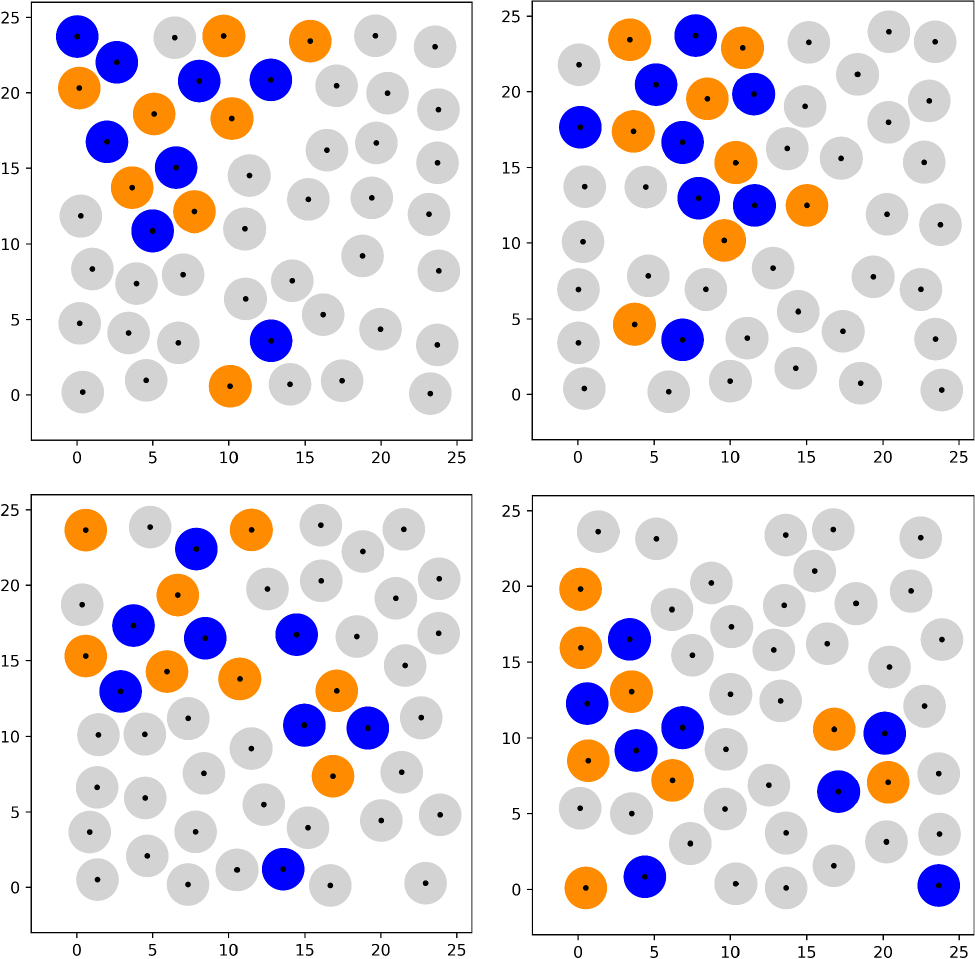}}
\caption{Membrane tension control of the clustering of inclusions with curvature $\pm c$ (blue and orange) mixed with inclusions with zero curvature (gray). The images are four typical snapshots of a Metropolis Monte Carlo simulation at large membrane tension, showing dynamic clusters with local antiferromagnetic order. Parameters: $c=0.5\,a^{-1}$, $\kappa=30\,\kT$ and $\bar\sigma=\sigma a^2/\kappa=0.16$.}
\label{fig:3species}
\end{figure}

Since real biological membranes are crowded with proteins, we considered also mixtures composed of oppositely curved proteins and cylindrically shaped proteins (imposing $c=0$). At the pairwise level, either with or without tension, curvature-inducing proteins repel cylindrical proteins, as can be seen from eqs.~\eqref{eq:Goulian} and \eqref{eq:Weikl} for $c_1c_2=0$. We found that the presence of cylindrical proteins promotes the formation of mixed clusters with local antiferromagnetic order (fig.~\ref{fig:3species}). These dynamic clusters, which form at $\bar\sigma\approx0.1$ for the parameter set~\eqref{eq:params}, disappear when decreasing the tension by a factor $10-100$.

\section{Curving inclusions subject to short-range interactions}
\begin{figure}
\centerline{\includegraphics[width=.9\columnwidth]{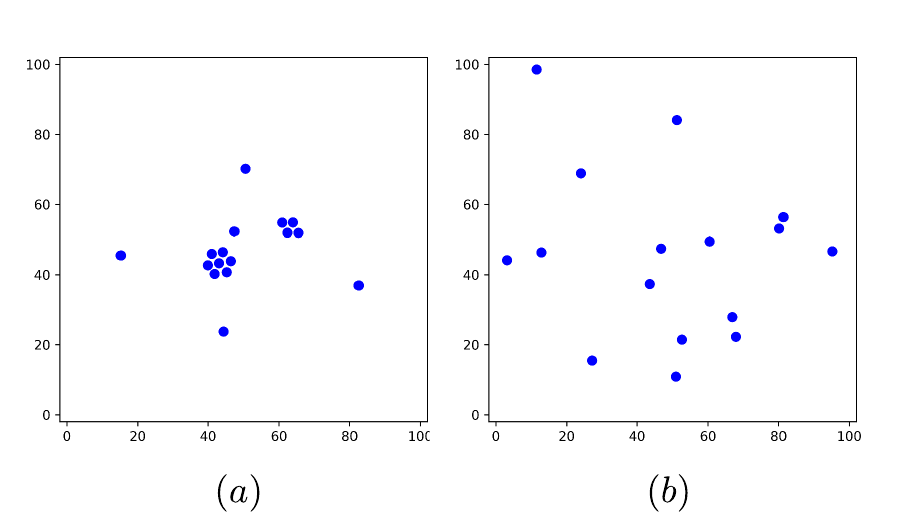}}
\caption{Membrane tension control of the aggregation of inclusions of curvature $c$ subject to specific short-range interactions of depth $e_0=-5\,\kT$ and range $\epsilon=0.5a$. Snapshots of a Metropolis Monte Carlo simulation (a) at low tension $\bar\sigma=\sigma a^2/\kappa<\bar\sigma_0$, (b) at large tension $\bar\sigma>\bar\sigma_0$. Parameters: $c=0.5\,a^{-1}$, $\kappa=30\,\kT$, $\bar\sigma_0=0.03$.}
\label{fig:specific}
\end{figure}

Membrane tension can also have the opposite effect of breaking clusters. Since real membrane proteins are often subject to specific short-range interactions significantly larger than $\kT$, e.g., hydrogen bonds, ionic/dipolar interactions, hydrophobic mismatch interactions~\cite{Jiang:2022}, we have considered curvature-inducing proteins of equal curvature $c$ subjected to a short-range potential well of depth $e_0=-5\,\kT$ acting at particle separations in the interval  $r\in[r_\text{min},r_\text{min}+\epsilon]$, with $r_\text{min}=3a$ and $\epsilon=0.5a$. {We found that the dynamic clusters that are formed at small tension become statistically smaller when the tension increases, and break up above a critical tension $\bar\sigma_0$ that depend on the parameters (fig.~\ref{fig:specific}).}

In conclusion, interesting effects can be expected from the interplay between membrane tension and multibody interactions. Tension makes the interactions short-ranged, but it reinforces them at short distances, where it really counts. Whereas curved inclusions repel each other whatever their sign at low tension, oppositely curved inclusions attract each other at large tension and tend to arrange antiferromagnetically, in a square lattice, due to their strong multibody interactions. We found also that the enhanced repulsion between identical curved proteins at large tension can break clusters stabilized by short-range forces, {which supports the results of Ref.~\cite{Weitz:2013} on cluster size distribution. We were not able to investigate, however, wether oppositely curved inclusions segregate into small clusters of opposite curvature at moderate tensions, as we could not simulate large enough equilibrium systems.} Our predictions could be investigated experimentally by monitoring the clustering of curvature-inducing proteins using FRET or grazing incidence X-ray diffraction.

\section{Appendix: Green function's derivatives}
In polar coordinates $\bm r=(x,y)=(r\cos\theta,r\sin\theta)$, with
\begin{align}
  G(\bm r)=-\frac{\xi^2}{2\pi}\left[K_0\!\left(\frac r\xi\right)+\ln\left(\frac r\xi\right)\right],
\end{align}
we obtain
\begin{widetext}
\begin{align}
\partial^4_xG(\bm r)&=\frac{24 \xi ^4 \cos (4 \theta )-3 r^4 \sin ^2(2 \theta ) K_0(r/\xi)-2 r^2
   K_2(r/\xi) \left[\cos (4 \theta ) \left(6 \xi ^2+r^2\right)+r^2 \cos (2 \theta
   )\right]}{8 \pi  \xi ^2 r^4 },
\\
\partial^2_x\partial^2_yG(\bm r)&=\frac{4 \cos (4 \theta ) \left[\left(r^4+6 \xi ^2 r^2\right) K_2(r/\xi)-12 \xi
   ^4\right]-r^4 \left[3 \cos (4 \theta )+1\right] K_0(r/\xi)}{16 \pi  \xi ^2 r^4 },
   \\
\partial^3_x\partial_yG(\bm r)&=-\frac{\sin (2 \theta ) \left\{r^2 K_2(r/\xi) \left[4 \cos (2 \theta ) \left(6 \xi
   ^2+r^2\right)+r^2\right]-3 \cos (2 \theta ) \left[16 \xi ^4+r^4 K_0(r/\xi)\right]\right\}}{8 \pi  \xi ^2 r^4 },
\\
\partial^4_yG(\bm r)&=\frac{24 \xi ^4 \cos (4 \theta )-3 r^4 \sin ^2(2 \theta ) K_0(r/\xi)-2 r^2
   K_2(r/\xi) \left[\cos (4 \theta ) \left(6 \xi ^2+r^2\right)-r^2 \cos (2 \theta
   )\right]}{8 \pi  \xi ^2 r^4 },
\\
\partial_x\partial^3_yG(\bm r)&=\frac{\sin (2 \theta ) \left\{r^2 K_2(r/\xi) \left[4 \cos (2 \theta ) \left(6 \xi
   ^2+r^2\right)-r^2\right]-3 \cos (2 \theta ) \left[16 \xi ^4+r^4 K_0(r/\xi)\right]\right\}}{8 \pi  \xi ^2 r^4 }.
\end{align}

These quantities are regularized in $r=0$ using the upper-wavevector cutoff $q_\text{max}=2/a$:
\begin{align}
\partial^n_x\partial^m_yG(\bm 0)=
\frac{\xi^2}{(2\pi)^2}\int_0^{q_\text{max}}\!dq\,\frac{q^3}{1+q^2\xi^2}
\int_0^{2\pi}\!d\theta\,\cos^n(\theta)\cos^m(\theta),
\end{align}
yielding the results after Eq.~\eqref{eq:matrix} in the main text.
\end{widetext}

\acknowledgments
I thank P. Bassereau, P. Galatola and F. van Wijland for fruitful discussions.

\end{document}